\documentclass[12pt]{iopart}

\usepackage{iopams}
\usepackage{bm,algorithm,algorithmic}
\usepackage{graphicx}
\usepackage{subfigure}
\usepackage{cite}
\usepackage[dcucite,abbr]{harvard}%
\begin{document}

\title[Cine-CBCT reconstruction]{Cine cone beam CT reconstruction using low-rank matrix factorization: algorithm and a proof-of-princple study}

\author{Jian-Feng Cai}
\address{Department of Mathematics, The University of Iowa, Iowa City, IA 52242, USA}
\author{Xun Jia}
\address{Center for Advanced Radiotherapy
Technologies and Department of Radiation Medicine and Applied Sciences, University of California
San Diego, La Jolla, CA 92037, USA}
\author{Hao Gao}
\address{Department of Mathematics,University of California Los Angeles, Los Angeles, CA 90095, USA}
\author{Steve B. Jiang}
\address{Center for Advanced Radiotherapy
Technologies and Department of Radiation Medicine and Applied Sciences, University of California
San Diego, La Jolla, CA 92037, USA}
\author{Zuowei Shen}
\address{Department of Mathematics, National University of Singapore, Singapore 119076}
\author{Hongkai Zhao}
\address{Department of Mathematics, University of California Irvine, Irvine, CA 92697, USA}

\newpage

\begin{abstract}
Respiration-correlated CBCT, commonly called 4DCBCT, provide respiratory phase-resolved CBCT images. In many clinical applications, it is more preferable to reconstruct true 4DCBCT with the 4th dimension being time, i.e., each CBCT image is reconstructed based on the corresponding instantaneous projection. We propose in this work a novel algorithm for the reconstruction of this truly time-resolved CBCT, called cine-CBCT, by effectively utilizing the underlying temporal coherence, such as periodicity or repetition, in those cine-CBCT images. Assuming each column of the matrix $\bm{U}$ represents a CBCT image to be reconstructed and the total number of columns is the same as the number of projections, the central idea of our algorithm is that the rank of $\bm{U}$ is much smaller than the number of projections and we can use a matrix factorization form $\bm{U}=\bm{L}\bm{R}$ for $\bm{U}$. The number of columns for the matrix $\bm{L}$ constraints the rank of $\bm{U}$ and hence implicitly imposing a temporal coherence condition among all the images in cine-CBCT. The desired image properties in $\bm{L}$ and the periodicity of the breathing pattern are achieved by penalizing the sparsity of the tight wavelet frame transform of $\bm{L}$ and that of the Fourier transform of $\bm{R}$, respectively. A split Bregman method is used to solve the problem. In this paper we focus on presenting this new algorithm and showing the proof of principle using simulation studies on an NCAT phantom.
\end{abstract}

\noindent{\it Keywords}: cine-CBCT, reconstruction, low-rank matrix factorization

\submitto{\PMB}
\maketitle

\section{Introduction}

When cone beam computed tomography (CBCT) is applied to thorax or upper abdomen regions, motion-induced artifacts, such as blurring or distortion, become a serious problem, because different x-ray projections correspond to different volumetric CBCT images due to patient respiratory motion. To overcome this problem, four-dimensional CBCT (4DCBCT)\cite{JSLZ:MP:2005,SKMM:PMB:2005,TLLX:MP:2006} has been developed. In such a modality, all x-ray projections are first retrospectively grouped into different respiratory phase bins according to a breathing signal tagged on every projection image. A set of CBCT images are then reconstructed, each at a breathing phase, under the assumption that the projections placed into each bin correspond to the same or similar CBCT image. The number of phase bins is usually empirically chosen by a user based on the consideration of balancing temporal resolution and image quality. On one hand, a high temporal resolution requires a large number of phase bins, which leads to insufficient number of projections available to each phase and hence degraded quality of reconstructed CBCT images. On the other, a small number of phase bins results in low temporal resolution, as well as a relatively large bin width and the associated residual motion artifacts in the reconstructed images. To maintain clinically acceptable temporal resolution and image quality, 4DCBCT acquisition protocols such as slow gantry rotations and multiple gantry rotations have been invented to increase the total number of projections\cite{TLLX:MP:2006,LGTM:MP:2007,LX:IJROBP:2007}.

Conventional 4DCBCT reconstruction approach reconstructs CBCT images at different phases independent of each other\cite{JSLZ:MP:2005}. This straightforward method neglects the temporal correlations of CBCT images at different phases. In contrast, a collaborative reconstruction scheme has been recently proposed\cite{JLDT:MICCAI:10,GCSZ:PMB:11,TJDLJ:MP:2011,ZC:PMB:2011,JTLSJ:MP:2012}. Among them, \citeasnoun{JLDT:MICCAI:10} and \citeasnoun{TJDLJ:MP:2011} proposed a temporal non-local means method to constraint that the reconstructed CBCT images at neighboring phases must contain repetitive anatomical features. The locations of these features are allowed to vary among phases.  \citeasnoun{ZC:PMB:2011} first reconstruct an average CBCT image using all projections at all phases and impose a similarity constraint between this average image and the CBCT at each phase.  These methods explicitly enforces the similarity among reconstructed CBCT images. \citeasnoun{GCSZ:PMB:11} utilizes robust PCA techniques for the 4DCBCT reconstruction problem. This method restores a matrix whose columns are the 4DCBCT images at all phases under the assumption that, after a proper transformation, this matrix can be decomposed as the sum of a low-rank matrix corresponding to an almost static background and a sparse matrix corresponding to a moving foreground. The low-rank constraint achieved by minimizing the nuclear norm \cite{CCS:SIOPT:10} of the associated matrix implicitly imposes the inter-phase similarity of the background. However, in the context of anatomical motion in thorax or upper abdomen, there is no clear distinction between foreground and background. Besides, as pointed out in \cite{CRPW:Arxiv:10}, the nuclear norm minimization always finds a matrix factorization with mutually orthogonal factors. In this sense, it may have some unfavourable bias on the resulting low-rank matrices.

Although commonly called 4DCBCT, respiration-correlated CBCT \cite{FMYL:MP:2003} is a more accurate name for this imaging technique, since its 4th dimension is actually the respiratory phase rather than time. Clinically, it is more preferable to reconstruct true 4DCBCT with the 4th dimension being time, i.e., one CBCT image is reconstructed based on the corresponding instantaneous projection, if possible. This truly time-resolved CBCT, called cine-CBCT in this work to avoid confusion, encounters an apparent technical challenge of reconstructing a CBCT image based on only one projection. Previously we have tried to overcome this challenge by utilizing prior 4DCT images of the same patient \cite{LJLGFMJ:MP:2010,LLJGFMSJ:MP:2011}. Particularly, we have built a lung motion model for the patient by performing a principle component analysis(PCA) on the motion vector fields obtained from deformable registration on prior 4DCT images. We discovered that only a few principle components are sufficient to represent the lung motion to a satisfactory degree of accuracy. The reconstruction of a volumetric image based on one acquired x-ray projection is then achieved by finding those PCA coefficients, so that the projection of the corresponding CBCT matches the acquired one. However, this method heavily depends on how similar the current CBCT images are to the prior 4DCT images and how accurately the PCA parameters can be determined using one projection.

In this paper, we will show the principle of proof of an innovative algorithm for cine-CBCT reconstruction without using any prior images. The basic idea is to  maximally exploit the temporal correlations, both local and nonlocal, of the patient anatomy at various time points, namely smooth variation of CBCT images at successive time points and the periodicity of all images during a scan, using recent mathematical developments. Specifically, a factorization of the matrix consisting of all the CBCT images into a product of two matrices is used to explicitly enforce a low-rank condition of the matrix that is implied by the local or nonlocal temporal image similarity. Moreover, sparsity conditions are enforced under appropriate transforms, i.e. wavelet transform in space for one matrix and Fourier transform in time for the other matrix, to satisfy desired properties for the sequence of CBCT images in both space and time. Simulation studies have demonstrated promising results for our method for this challenging problem.

\section{Methods and Materials}\label{SecProposed}

\subsection{Model}
In this proof-of-principle study, we consider the reconstruction of a 2D slice of the CBCT  to illustrate the principles of the new algorithm. The basic ideas, however, can be easily generalized into 3D contexts. Let the unknown 2D image be a function $u$ defined in $\mathbb{R}^2$. When the X-ray source is placed with angle $\theta$, the projection measured by the imager at location $z$ is
\begin{equation}\label{EqRadon}
f_{\theta}(z)=\mathcal{P}_{\theta}u(z):=\int_{0}^{\ell_z}u(\bm{x}_{\theta}+s\bm{r}_z)ds
\end{equation}
where $\bm{x}_{\theta}\in\mathbb{R}^2$ is the coordinate of the X-ray source, and $\bm{r}_z\in\mathbb{R}^{2}$ and $\ell_z$ are respectively the direction and the length of the line connecting the X-ray source and the location $z$ on the imager. The operator $\mathcal{P}$ in \eref{EqRadon} is also known as Radon transform. If $f_{\theta}(z)$ is sampled with respect to $z$, the resulting projection data can be written as a vector $\bm{f}_{\theta}\in\mathbb{R}^{M}$ that obeys
\begin{equation}\label{EqCT1P}
\bm{P}_{\theta}\bm{u}=\bm{f}_{\theta}.
\end{equation}
Here $\bm{P}_{\theta}\in\mathbb{R}^{M\times N}$ and $\bm{u}\in\mathbb{R}^{N}$ are the discretization of $\mathcal{P}_{\theta}$ and $u$ in \eref{EqRadon} respectively. Suppose that we have $T$ projections where the X-ray source is placed with angels $\theta_1,\theta_2,\ldots,\theta_T$ respectively. Let $\bm{P}_i$ and $\bm{f}_i$ stand for $\bm{P}_{\theta_i}$ and $\bm{f}_{\theta_i}$ respectively. Then, by putting (\ref{EqCT1P}) with different angles together, the CT projection can be written into a system of linear equations
\begin{equation}\label{EqCT}
\bm{P}\bm{u}=\bm{f},
\end{equation}
where $\bm{P}=[\bm{P}_{1};~\bm{P}_{2};~\ldots;~\bm{P}_{T}]\in\mathbb{R}^{MT\times N}$ and $\bm{f}=[\bm{f}_{1};~\bm{f}_{2};~\ldots;~\bm{f}_{T}]\in\mathbb{R}^{MT}$. In other words, the CT image reconstruction problem is to recover image $\bm{u}$ from its partial Radon transform.

In the context of cine-CBCT reconstruction, instead of only one unknown image $u$ as in the CBCT problem, there is a set of unknown images, each associated with a projection, denoted by $\{\bm{u}_{1},\bm{u}_{2},\ldots,\bm{u}_{T}\}$ and the projection condition
in \eref{EqCT} may be modified to
\begin{equation}\label{Eq4DCT}
\bm{P}_{i}\bm{u}_{i}=\bm{f}_{i},\quad i=1,2,\ldots,T.
\end{equation}
If we write all the unknown images in a matrix form
$\bm{U}=[\bm{u}_{1},\bm{u}_{2},\ldots,\bm{u}_{T}]$, \eref{Eq4DCT} can be rewritten into a compact matrix equation
\begin{equation}\label{Eq4DCTMatrix}
\mathcal{P}\bm{U}=\bm{F}
\end{equation}
where $\mathcal{P}\bm{U}=[\bm{P}_{1}\bm{u}_{1},\ldots,\bm{P}_{T}\bm{u}_{T}]$ and $\bm{F}=[\bm{f}_{1},\ldots,\bm{f}_{T}]$.

Because of the temporal coherence among the unknown images, it is expected that the columns of the matrix $\bm{U}$ formulated as such, namely the CBCT images, are almost linear dependent, and hence the rank of the resulting matrix $\bm{U}$ is very low, much smaller than the number of probections. With the low rank assumption, the number of intrinsic unknowns of \eref{Eq4DCTMatrix} are possibly less than the number of measurements, and therefore it is possible to reconstruct $T$ CBCT images in $\bm{U}$ from $T$ projections. To incorporate the underlying low-rank assumption into the cine-CBCT reconstruction process, we would like to explore matrix factorization based low-rank models in this study. More specifically, we enforce a decomposition form of the unknown matrix $\bm{U}$ of images as $\bm{U}=\bm{L}\bm{R}$ where $\bm{L}\in\mathbb{R}^{N\times K}$ and $\bm{R}\in\mathbb{R}^{K\times T}$ for a small integer $K$. From basic linear algebra knowledge, any matrix $\bm{U}$ of rank $K$ can be represented in $\bm{U}=\bm{L}\bm{R}$ and conversely, the rank of $\bm{U}$ is at most $K$ given the factorization form. In our algorithm, the value of $K$ is specified by the user and our method regarding the selection of $K$ will be presented later. We would like to point out that there is another popular approach for imposing the low rank condition that penalizes the nuclear norm of the matrix $\bm{U}$, namely the sum of the singular values. Compare these two methods, the latter always finds orthogonal basis because of the involved singular value decomposition process. In contrast, our factorization method does not require this implicit orthogonality and hence attains the advantage of avoiding unfavourable bias on the resulting low-rank matrices \cite{CRPW:Arxiv:10}. In some sense, the approach of penalizing the nuclear norm of $\bm{U}$ is like Principle Component Analysis and our approach is like Independent Component Analysis. As a consequence, each of our basis, i.e., each column of $\bm{L}$, represents more meaningful images and hence we can enforce some desired image properties on $\bm{L}$, such as sparsity under wavelet transform.

This matrix factorization approach also allows for the additional regularizations on $\bm{L}$ and $\bm{R}$, so that $\bm{U}=\bm{L}\bm{R}$ carries desirable physical properties. First, the columns of the matrix $\bm{L}$, corresponding to  images, can be interpreted as a basis that efficiently represents all the columns of $\bm{U}$. Notice that images usually have sparse coefficients under suitable transforms such as wavelet tight frames \cite{RS:JFA:97}. Let $\mathcal{D}$ be such a transform. Since the columns of $\bm{L}$ are images, we want $\mathcal{D}\bm{L}$ to be sparse. It is well-known that $\ell_1$-norm minimization leads to sparse solutions. Therefore, the $\ell_1$-norm $\|\mathcal{D}\bm{L}\|_1$ is penalized.
%Since images usually have sparse representations under suitable transformations, we can penalize %$\|\mathcal{D}\bm{L}\|_1$, where $\mathcal{D}$ is a tight wavelet frame \cite{RS:JFA:97}
%%\textbf{Commenst: is this what you used, Jianfeng? please cite}
%sparsifying transformation applied to each column of $\bm{L}$, to encourage the columns of $\bm{L}$ to be an image.
Second, the rows of $\bm{R}$ are the coefficients of $\bm{U}$ under the basis $\bm{L}$. Since the respiratory motion is near periodic, each row of $\bm{R}$ can be modeled as a period signal in time. In other words, the Fourier transform of each row of $\bm{R}$ contains very few nonzero entries. Since the $\ell_1$-norm minimization helps get sparse solutions, we can use $\|\mathcal{F}\bm{R}\|_1$ as a regularization term for $\bm{R}$, where $\mathcal{F}$ is the Fourier transform applied to each row.

Altogether, we propose to reconstruct the cine-CBCT by solving an optimization problem
\begin{equation}\label{Eq4DCTLRnoiseless}
\min_{\bm{L},\bm{R}}\|\mathcal{D}\bm{L}\|_1+\lambda\|\mathcal{F}\bm{R}\|_1,
\quad
\mbox{s.t. }\mathcal{P}(\bm{L}\bm{R})=\bm{F},
\end{equation}
where $\lambda$ is a parameter that balances the sparsity of $\mathcal{D}\bm{L}$ and $\mathcal{F}\bm{R}$. In practice, \eref{Eq4DCTMatrix} is barely satisfied, because there is always unavoidable noise in the measurements $\bm{F}$. Moreover, in reality the matrix $\bm{U}$ is only approximately low-rank and explicitly enforcing a low-rank representation leads to error. For these considerations, we solve
\begin{equation}\label{Eq4DCTLR}
\min_{\bm{L},\bm{R}}\|\mathcal{D}\bm{L}\|_1+\lambda\|\mathcal{F}\bm{R}\|_1,
\quad
\mbox{s.t. }\|\mathcal{P}(\bm{L}\bm{R})-\bm{F}\|_F^2\leq\sigma^2,
\end{equation}
where $\|\cdot\|_F$ is the Frobenius norm, namely $\|\bm{A}\|_F=\sqrt{\sum_{i,j}\bm{A}_{i,j}^2}$ for a matrix $\bm{A}$.
In \eref{Eq4DCTLR}, $\lambda$ is again a parameter balancing the sparsity of $\mathcal{D}\bm{L}$ and $\mathcal{F}\bm{R}$, and $\sigma$ is a parameter to control to what extent the violation of \eref{Eq4DCTMatrix} is allowed.

\subsection{Algorithm}
Let us first consider the algorithm for solving Equation \eref{Eq4DCTLRnoiseless}. We use a split Bregman method \cite{GO:SIIMS:09,COS:MMS:09} (also known as augmented Lagrangian method) to solve this problem. In particular, by introducing some auxiliary variables $\bm{C}$ and $\bm{D}$, the problem in \eref{Eq4DCTLRnoiseless} is equivalent to
\begin{equation}\label{EqAL}
\min_{\bm{C},\bm{D},\bm{L},\bm{R}} \|\bm{C}\|_1 +\lambda\|\bm{D}\|_1,
\quad\mbox{s.t. }\mathcal{P}(\bm{L}\bm{R})=\bm{F},~\bm{C}=\mathcal{D}\bm{L},
~\bm{D}=\mathcal{F}\bm{R}.
\end{equation}
The augmented Lagrangian is
\begin{eqnarray*}
\fl E(\bm{C},\bm{D},\bm{L},\bm{R},\bm{Z},\bm{Z}_1,\bm{Z}_2)=
\|\bm{C}\|_1 +\lambda\|\bm{D}\|_1 +
\langle\bm{Z},\mathcal{P}(\bm{L}\bm{R})-\bm{F}\rangle
+\frac{\mu}{2}\|\mathcal{P}(\bm{L}\bm{R})-\bm{F}\|_F^2\\
+\langle\bm{Z}_1,\bm{C}-\mathcal{D}\bm{L}\rangle
+\frac{\mu_1}{2}\|\bm{C}-\mathcal{D}\bm{L}\|_F^2
+\langle\bm{Z}_2,\bm{D}-\mathcal{F}\bm{R}\rangle
+\frac{\mu_2}{2}\|\bm{D}-\mathcal{F}\bm{R}\|_F^2,
\end{eqnarray*}
where $\langle\cdot,\cdot\rangle$ is the inner product, and $\bm{Z}$, $\bm{Z}_1$, and $\bm{Z}_2$ are Lagrange multipliers.
%As \eqref{EqOpt2} is not a convex optimization, there is a duality gap between the primary problem \eqref{EqOpt2} and the dual problem
%$$
%\max_{\bm{Z_1},\bm{Z_2}}\min_{\bm{D},\bm{L},\bm{R}}E(\bm{D},\bm{L},\bm{R},\bm{Z}_1,\bm{Z}_2).
%$$
With appropriate fixed $\bm{Z}$, $\bm{Z}_1$, and $\bm{Z}_2$, an optimal $\bm{C},\bm{D},\bm{L},\bm{R}$ can be found by simply minimizing $E(\bm{C},\bm{D},\bm{L},\bm{R},\bm{Z},\bm{Z}_1,\bm{Z}_2)$ with respect to $(\bm{C},\bm{D},\bm{L},\bm{R})$. Therefore, the trick is to determine $\bm{Z}$, $\bm{Z}_1$, and $\bm{Z}_2$. In the augmented Lagrangian algorithm, we use
\begin{equation}\label{EqALA}
\cases
{(\bm{C}^{(k+1)},\bm{D}^{(k+1)},\bm{L}^{(k+1)},\bm{R}^{(k+1)})=\arg\min_{\bm{C},\bm{D},\bm{L},\bm{R}}E(\bm{C},\bm{D},\bm{L},\bm{R},\bm{Z}^{(k)},\bm{Z}_1^{(k)},\bm{Z}_2^{(k)}),\\
\bm{Z}^{(k+1)}=\bm{Z}^{(k)}+(\mathcal{P}(\bm{L}^{(k+1)}\bm{R}^{(k+1)})-\bm{F}),\\
\bm{Z}_1^{(k+1)}=\bm{Z}_1^{(k)}+(\bm{C}^{(k+1)}-\mathcal{D}\bm{L}^{(k+1)}),\\
\bm{Z}_2^{(k+1)}=\bm{Z}_2^{(k)}+(\bm{D}^{(k+1)}-\mathcal{F}\bm{R}^{(k+1)}).
}
\end{equation}
As for the first minimization problem in \eref{EqALA}, we use an alternative minimization algorithm. As such, we minimize $L$ sequentially in four steps, each with respect to one of the variables $\bm{C}$, $\bm{D}$, $\bm{L}$, and $\bm{R}$ at a time, while keep the rest fixed. These four subproblems can be solved easily by either entry-wise soft-thresholding (for $\bm{C}$ and $\bm{D}$) and linear equation solvers (for $\bm{L}$ and $\bm{R}$), e.g. conjugate gradient method. The algorithm is summarized in Algorithm \ref{Alg1}, where $\mathcal{T}$ is the soft-thresholding operator defined by $[\mathcal{T}_{\alpha}(\bm{A})]_{ij}=\mbox{sign}([\bm{A}]_{ij})\cdot\max\{|[\bm{A}]_{ij}|-\alpha,0\}$.

%which are solved as follows
%\begin{itemize}
%\item The solution to the minimization of $L$ with respect to $\bm{C}$ or $\bm{D}$ is equivalent to a soft-thresholding operation, which can be done straightforwardly.
%%\textbf{Comments: we need write explicitly the soft threshholding operation}
%\
%
%\item The solution to the minimization of $L$ with respect to $\bm{L}$ or $\bm{R}$ is the solution to a least-squares problem, which can be solved by a linear equation solver such as the conjugate gradient method.
%\end{itemize}
%
%\textbf{Comments: A table summarizing the algorithm would be helpful.}

\begin{algorithm}[h!]
\caption{}\label{Alg1}
\begin{algorithmic}[1]
\STATE Solve the first subproblem in \eref{EqALA} by alternative minimization iteratively as follows
\begin{enumerate}
\item Update $\bm{C}\leftarrow \mathcal{T}_{1/\mu_1}(\mathcal{D}\bm{L}-\bm{Z}_1^{(k)}/\mu_1)$.
\item Update $\bm{D}\leftarrow \mathcal{T}_{\lambda/\mu_2}(\mathcal{F}\bm{R}-\bm{Z}_2^{(k)}/\mu_2)$.
\item Update $\bm{L}\leftarrow \arg\min_{\bm{L}}\frac{\mu}{2}\|\mathcal{P}(\bm{L}\bm{R})-\bm{F}+
\bm{Z}^{(k)}/\mu\|_F^2+\frac{\mu_1}{2}\|\mathcal{D}\bm{L}-\bm{C}-\bm{Z}_1^{(k)}/\mu_1\|_F^2$.
\item Update $\bm{R}\leftarrow \arg\min_{\bm{R}}\frac{\mu}{2}\|\mathcal{P}(\bm{L}\bm{R})-\bm{F}+
\bm{Z}^{(k)}/\mu\|_F^2+\frac{\mu_2}{2}\|\mathcal{F}\bm{R}-\bm{D}-\bm{Z}_2^{(k)}/\mu_2\|_F^2$.
\end{enumerate}
\STATE If $\|\mathcal{P}(\bm{L}^{(k+1)}\bm{R}^{(k+1)})-\bm{F}\|_F^2$ is small enough, then return.
\STATE $\bm{Z}^{(k+1)}=\bm{Z}^{(k)}+(\mathcal{P}(\bm{L}^{(k+1)}\bm{R}^{(k+1)})-\bm{F})$.
\STATE $\bm{Z}_1^{(k+1)}=\bm{Z}_1^{(k)}+(\bm{C}^{(k+1)}-\mathcal{D}\bm{L}^{(k+1)})$.
\STATE $\bm{Z}_2^{(k+1)}=\bm{Z}_2^{(k)}+(\bm{D}^{(k+1)}-\mathcal{F}\bm{R}^{(k+1)})$.
\STATE $k=k+1$ and goto Step 1.
\end{algorithmic}
\end{algorithm}

One complexity of this problem comes from the non-convexity of the objective function in \eref{Eq4DCTLRnoiseless}. Since there may exist local minima, we have to make sure our algorithm get a desired solution. Yet, in our numerical experiments, this non-convexity issue is not found to be a problem for the following two reasons. First of all, our algorithm may find the global minimum of \eref{Eq4DCTLRnoiseless}. As pointed out in \citeasnoun{BM:MP:05}, the augmented Lagrangian algorithm attains the global minimum for non-convex objectives in low-rank factorization form under suitable assumptions. Similar algorithms to ours has been used in, e.g., SDPLR (Semi-definite programming via low rank factorization) \cite{BM:MP:03}. Secondly, we have also chosen the initial guess to the iterative algorithm in the Algorithm \ref{Alg1} carefully as following.
%an initial guess $\bm{L}^{(0)}$ and $\bm{R}^{(0)}$ such that $\bm{L}^{(0)}\bm{R}^{(0)}$ %is an (approximate) low rank solution to \eref{Eq4DCTMatrix}. Then, the solution by %\eref{EqALA} is refine  \eref{Eq4DCTLRnoiseless}.
We first solve a convex minimization problem
\begin{equation}\label{Eq:Nuc}
\min_{\bm{U}}~\frac12\|\mathcal{P}\bm{U}-\bm{F}\|_F^2+\lambda\|\bm{U}\|_*,
\end{equation}
where $\|\bm{U}\|_*$ is the nuclear norm, i.e., the summation of the singular values of $\bm{U}$. The nuclear norm minimization is able to find a lowest-rank solution $\bm{U}_*$ of \eref{Eq4DCTMatrix} within a precision \cite{CCS:SIOPT:10}. Then, the initial guess in the Algorithm \ref{Alg1} is chosen to be the best rank-$K$ approximation to $\bm{U}_*$. Let $\bm{U}_*=\bm{W}\bm{\Sigma}\bm{V}^T$ be a singular value decomposition. If the rank of $\bm{U}_*$ is less than $K$, we choose $\bm{L}^{(0)}=\bm{W}\bm{\Sigma}^{1/2}$ and $\bm{R}^{(0)}=\bm{\Sigma}^{1/2}\bm{V}^T$ as the initial guesses. If the rank of $\bm{U}_*$ exceeds $K$, we choose $\bm{L}^{(0)}=\bm{W}_K\bm{\Sigma}_K^{1/2}$ and $\bm{R}^{(0)}=\bm{\Sigma}_K^{1/2}\bm{V}^T_K$ as the initial guesses, where $\bm{W}_K,\bm{V}_K$ are the first $K$ columns of $\bm{W}$ and $\bm{V}$ respectively, and $\bm{\Sigma}_K$ is the $K\times K$ principle submatrix of $\bm{\Sigma}$. By this way, $\bm{L}^{(0)}\bm{R}^{(0)}$ is the best rank-$K$ approximation of $\bm{U}_*$.

As for the problem in \eref{Eq4DCTLR}, we still use the Algorithm \ref{Alg1}, but we stop the iteration as soon as $\|\mathcal{P}(\bm{L}\bm{R})-\bm{F}\|_F^2\leq\sigma^2$. By this way, we can get a quite good approximate solution to \eref{EqALA}. This approach has been previously used and discussed in similar mathematical problems but in other contexts, e.g., \cite{OBGXY:MMS:05,COS:MMS:09,COS:SIIMS:09}.

\subsection{Choice of $K$}
One practical issue in our algorithm is the selection of the parameter $K$ to control the matrix rank. In practice, we achieve this goal by the following procedure.

We first set $K$ to be a large enough number, e.g. $K=20$, and run our algorithm. Recall that the columns of $\bm{L}$ forms a basis to represent the images in $\bm{U}$. Because of the sufficiently large $K$ value in this trial run, the algorithm is forced to generate a set of $K$ basis vectors containing those good ones for $\bm{U}$, as well as the unnecessary bad ones. This is reflected by the fact that some columns of $\bm{L}$ in the solution make significant contributions to $\bm{U}$, while the other columns contribute little. For those unnecessary columns in $\bm{L}$, their presence introduces a small but observable amount of signals into the finally reconstructed images in $\bm{U}$ and compromises its quality. It is therefore desirable to eliminate them from the first place by setting the value of $K$ to be the number of the significant columns. A second reconstruction is then performed with this properly chosen $K$ value.

It remains to determine which columns are significant and which ones are negligible. Since the contribution of the $i$-th column of $\bm{L}$ to $\bm{U}$ is $\bm{L}(:,i)\bm{R}(i,:)$, any norm of  this project reflects the importance of this basis vector $\bm{L}(:,i)$. The larger norm is, the more significant $\bm{L}(:,i)$ is. In the experiments, we use the $\infty$-norm to identify the significance, which is the maximum of row absolute sums of a matrix. The column $\bm{L}(:,i)$ is considered to be negligible, if $\|\bm{L}(:,i)\bm{R}(i,:)\|_{\infty}$ is close to zero, and significant otherwise.

\subsection{Experiments}
\begin{figure}
\begin{center}
\includegraphics[width=.66\textwidth]{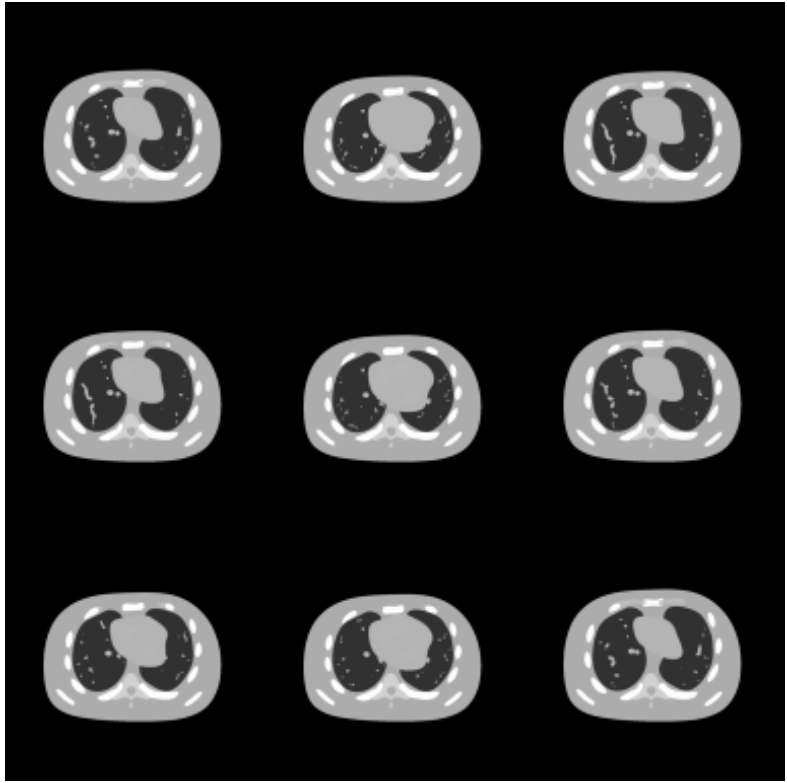}
\caption{The ground truth cine-CBCT images. From left to right, top to bottom: frame $40,80,\ldots,360$.}\label{fig0}
\end{center}
\end{figure}

We tested our cine-CBCT reconstruction algorithm on a digital NURBS-based cardiac-torso (NCAT) phantom \cite{SMBF:MP:08}, which generates a patient body in thorax region with detailed anatomical features and a realistic motion pattern. The patient respiratory period is 4 seconds. The CBCT gantry rotates about the patient at a constant speed for a full rotation in 59 seconds, in which 360 X-ray projections are acquired. At each X-ray projection acquisition, we compute the NCAT phantom image at the specific time point with a resolution of $128\times 128$, and the projection is then computed using a ray-tracing algorithm at the associated projection angle with a detector resolution of 256 bins. The patient breathing period and the gantry rotation period are chosen to be incommensurate deliberately, so that all the 360 patient images are distinct, although some of them visually look close. Some of the underlying true images are shown in Figure \ref{fig0}. Under this setup, the size of $\bm{U}$ is $16384\times 360$, and the size of $\bm{F}$ is $256\times 360$.

Two experiments are performed in this feasibility study, namely with and without noise signals in the measured projections $\bm{F}$. For the case with noise, a realistic CBCT projection noise signal corresponding to 0.5 mAs/projection level is generated according to a noise model \cite{WLLE:PMB:2008} and is added to the noise-free measurements $\bm{F}$. In each case, we will first present the results for the selection of the parameter $K$ and then the reconstruction results. Apart from visual inspection of the reconstructed cine-CBCT images, we have also quantitatively assessed the restored image quality using relative error as a metric defined by $\|\bm{L}\bm{R}-\bm{U}_{true}\|_F/\|\bm{U}_{true}\|_F$, where $\bm{L},\bm{R}$ are the outputs of the proposed method, and $\bm{U}_{true}$ is the matrix consisting of ground truth NCAT images.

\section{Experimental Results}\label{SecExperiments}

\subsection{Noiseless data}
\begin{figure}
\begin{center}
\includegraphics[width=.66\textwidth]{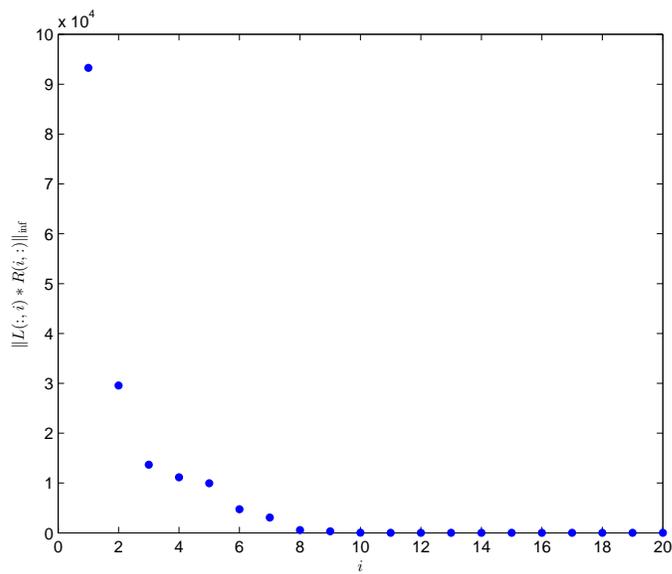}
\caption{$\|\bm{L}(:,i)\bm{R}(i,:)\|_{\infty}$ in descending order. The result is produced with $K=20$.}\label{figrank}
\end{center}
\end{figure}

\begin{figure}
\begin{center}
    {\includegraphics[width=.66\textwidth]{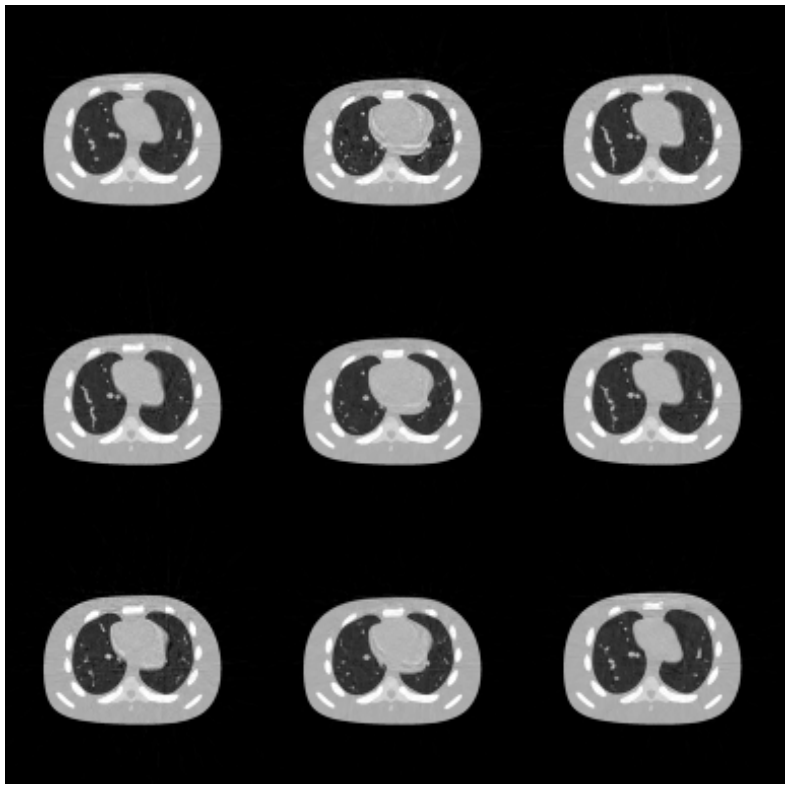}}
\caption{The restored cine-CBCT image from noiseless data. The relative error is $3.97\%$. From left to right, top to bottom: frame $40,80,\ldots,360$.}\label{fig1}
\end{center}
\end{figure}

\begin{figure}
\begin{center}
  \subfigure[The columns of $\bm{L}$. Each column is reshaped into an image.]%
    {\includegraphics[width=.88\textwidth]{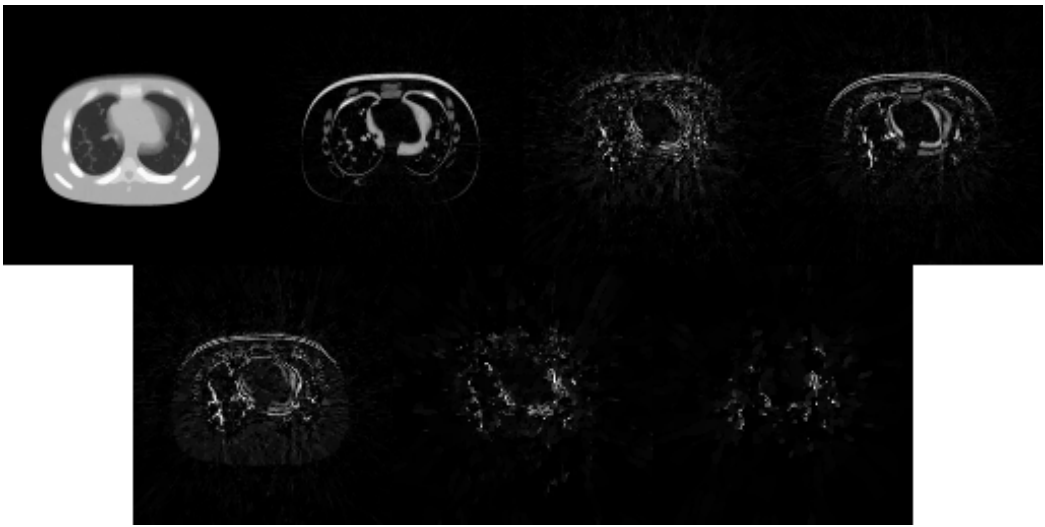}}
  \subfigure[The rows of $\bm{R}$.]%
    {\includegraphics[width=.88\textwidth]{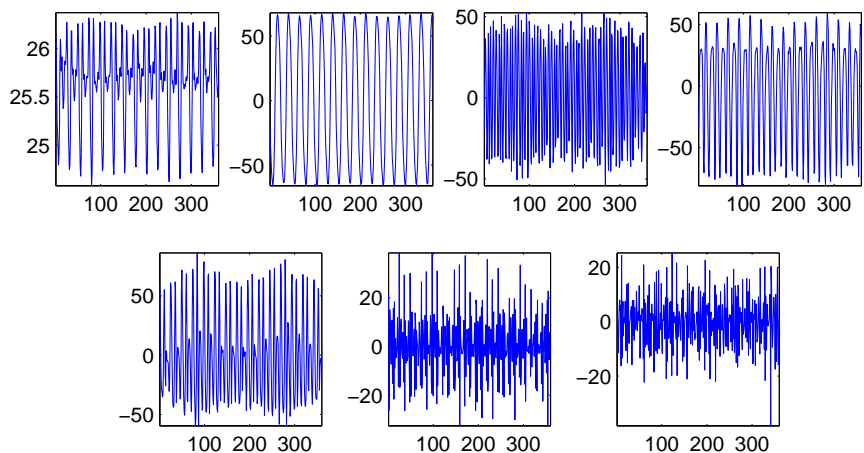}}
\end{center}
\caption{$\bm{L}$ and $\bm{R}$ in the cine-CBCT image reconstruction from noiseless data.}\label{fig11}
\end{figure}

For this case, the plot of $\|\bm{L}(:,i)\bm{R}(i,:)\|_{\infty}$ in a descending order as a function of the column index $i$ is first depicted in \Fref{figrank}, with a trial run of $K=20$.  Clearly, the 8th column and beyond contribute little to the reconstructed images and hence should be removed. As such,  $K=7$ is selected and the corresponding results are shown in \Fref{fig1} and \Fref{fig11}.
Comparing the restored images in \Fref{fig1} with the corresponding ground truth images in Figure \ref{fig0}, it is found that our algorithm is able to capture the motions of the anatomy and restore the structures, even those small ones inside the lung. Meanwhile, observable artifacts inside the heart and at its boundary also exist.  Quantitatively, the relative error of the restored cine-CBCT images is $3.97\%$.

To further look into the reconstruction results, we plot the columns of $\bm{L}$ in \Fref{fig11}(a), where each column is reshaped into a $128\times 128$ image. These images form the basis to represent all the reconstructed cine-CBCT images. It is observed that the first one is similar to an image averaged over all the cine-CBCT images. Its presence provides an overall structure that is common to all the images in the cine-CBCT. Meanwhile, other basis images represent differences between images of $\bm{U}$. We also plot the corresponding coefficients in $\bm{R}$ in \Fref{fig11}(b), which attain a periodically variation pattern, indicating the patient respiratory motion.

\subsection{Noisy data}

\begin{figure}
\begin{center}
\includegraphics[width=.66\textwidth]{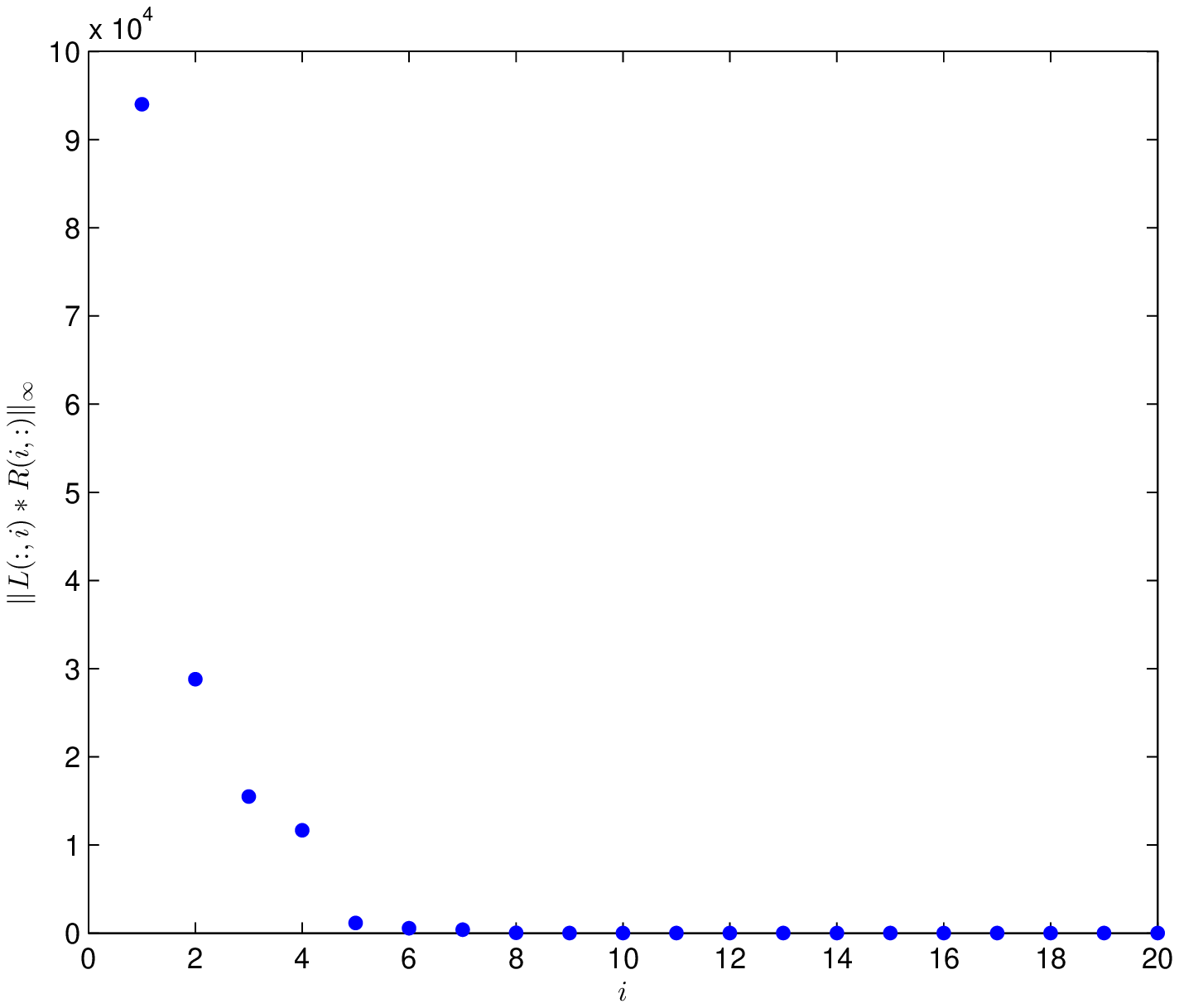}
\caption{$\|\bm{L}(:,i)\bm{R}(i,:)\|_{\infty}$ in descending order. The result is produced with $K=20$ and noise at 0.5 mAs/projection.}\label{figrank2}
\end{center}
\end{figure}

\begin{figure}
\begin{center}
  \subfigure[The restored cine-CBCT image with the relative error  $6.81\%$. From left to right, top to bottom: frame $40,80,\ldots,360$.]%
    {\includegraphics[width=.66\textwidth]{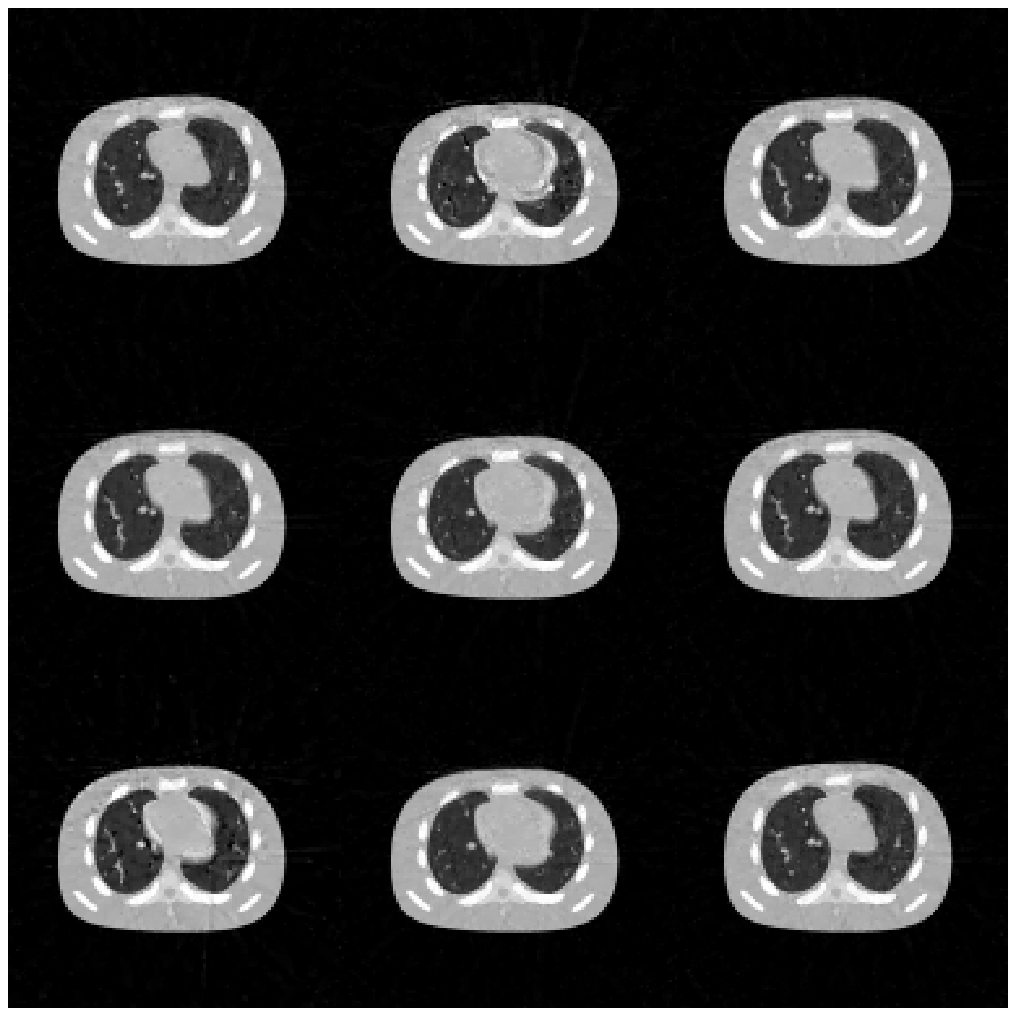}}\\
  \subfigure[The columns of $\bm{L}$. Each column is reshaped into an image.]%
    {\includegraphics[width=.88\textwidth]{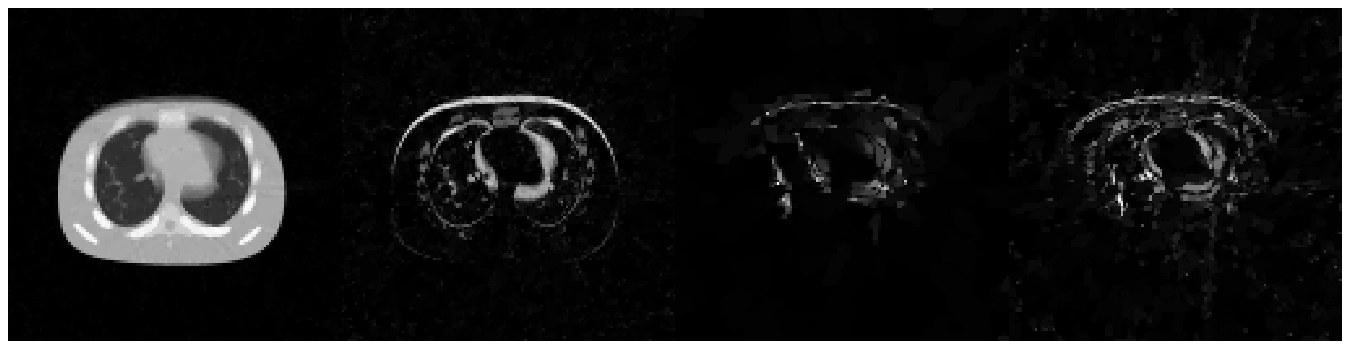}}\\
  \subfigure[The rows of $\bm{R}$.]%
    {\includegraphics[width=.99\textwidth]{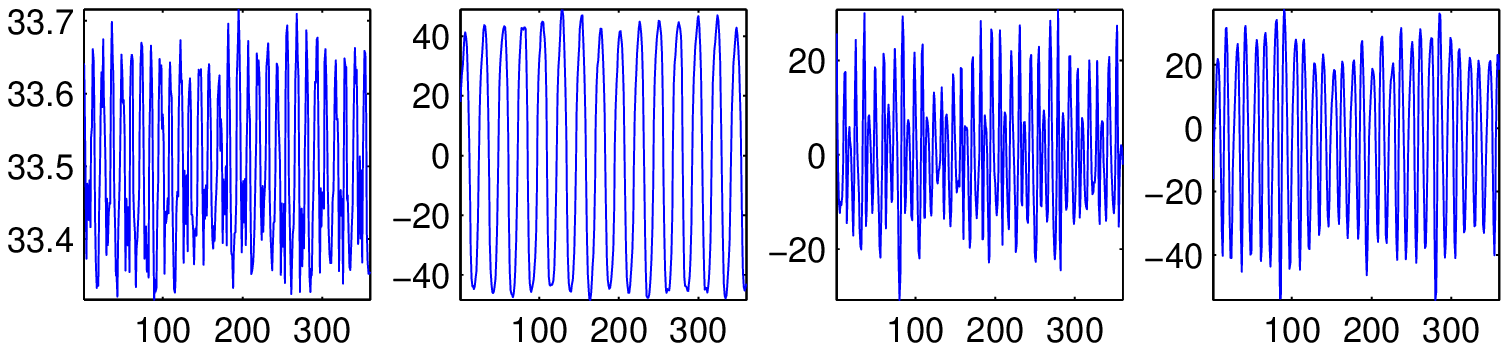}}
\end{center}
\caption{Result of cine-CBCT image reconstruction with noise at 0.5 mAs/projection.}\label{fig2}
\end{figure}

For the case with noise in the projection data, the parameter $K$ is selected as $K=4$ according to \Fref{figrank2}. \Fref{fig2}(a) depicts the reconstructed images. The relative error of the restored cine-CBCT images is $6.81\%$. The columns of $\bm{L}$ and the rows of $\bm{R}$ are plotted in \Fref{fig2}(b) and (c), respectively. Again, we see  that the columns of $\bm{L}$ are meaningful basis corresponding to the average image as well as the variations between them, and the rows of $\bm{R}$ oscillate periodically.

\section{Conclusion and Discussions}

In this work, we have proposed an innovative algorithm to reconstruct cine-CBCT to provide a set of time-dependent images in the thorax region. In contrast to the currently used 4DCBCT, where only respiratory phase-resolved imaging is achieved, the cine-CBCT modality offers a much higher temporal resolution, as the instantaneous patient anatomy based on a CBCT projection is retrieved. Cine-CBCT reconstruction is apparently a challenging problem due to the very limited projection information associated to each image to be reconstructed. Yet, by effectively incorporating the underlying temporal constraints satisfied by those cine-CBCT images into the reconstruction process, we demonstrate the feasibility of this modality. In particular, we assume a matrix factorization form $\bm{U}=\bm{L}\bm{R}$ for the matrixe $\bm{U}$ whose columns are the images to be reconstructed. The dimension of the matrices $\bm{L}$ constraints the rank of $\bm{U}$ and hence implicitly imposing a similarity condition among all the images in cine-CBCT. Moreover, near periodicity of the breathing pattern is reflected in $\bm{R}$, which is achieved by minimizing the sparsity of its Fourier coefficients. Simulation studies on an NCAT phantom serve as a very preliminary test to demonstrate the feasibility of our approach. The relative error for the reconstructed images is $3.97\%$ for the noise-free case and  $6.81\%$ for the case with noise.

In addition to the apparent advantage of improved temporal resolution in the cine-CBCT over 4DCBCT, the reconstruction of cine-CBCT is not as demanding regarding the amount of projection data as 4DCBCT, at least based on our simulation studies. Because of the phase binning in 4DCBCT, a certain number of projections are required in each bin to yield a CBCT image for the bin with a satisfactory quality. Hence, it is usually challenging to get a 4DCBCT with only one-minute scan and in practice, protocols of multiple rotations or a slow rotation is utilized. In contrast, the cine-CBCT effectively incorporates the inter-image correlations into the reconstruction and a set of images of acceptable quality are obtained even with a single-rotation scan protocol, as demonstrated in our studies.

Besides the modality of cine-CBCT, another potential application of our approach is the patient breathing signal extraction based on projection images. The respiratory signal is important for many clinical applications, such as projection sorting for 4DCBCT reconstruction. Currently, this signal is usually obtained by monitoring a surrogate known to move in synchronization with respiratory motion, e.g. patient surface or diaphragm. However, these surrogates may not be readily available for a particular patient and there also may exist phase shift between the motions of the lung and that of the surrogate. Therefore, it is desired to obtain a robust breathing signal from the projections itself. As pointed in our algorithm, the matrix entries of the matrix $\bm{R}$ contains the breathing information, as evidenced by Figure \ref{fig1}(c) and Figure \ref{fig2}(c), especially for those signals with a relatively large amplitude. Hence, our algorithm provides an effective method to extract the breathing signals from only projection information.

Despite the success in this feasibility study, there are a few practical issues as for the clinical applications of cine-CBCT. First, only one slice of the patient anatomy is reconstructed in this proof-of-principle study. When cone-beam geometry is considered to reconstruct volumetric CBCT images, although the fundamental principles in our algorithm can be easily applied, there potentially come some problems due to the extremely large data size involved. The computational efficiency will also be reduced. This problem may be resolved or alleviated by using more powerful computational platforms, such as computer graphic processing unit (GPU) \cite{JLS:MP:2010}. Yet, the algorithm implementation and optimization are not straightforward for the practical considerations of the limited memory and the non-trivial parallel computation schemes in many steps. For instance, solving the least square problems in the Algorithm \ref{Alg1} requires carefully implementations of matrix multiplications and the associated matrix-vector operations. These issues require further investigations. Another issue that may potentially deteriorate the image quality is the irregularity of breathing pattern. The penalty term regarding the sparsity of the Fourier coefficients in \eref{Eq4DCTLRnoiseless} favors those breathing patterns with a high level of periodicity. While this approach effectively imposes the periodicity aspect of the temporal correlation among cine-CBCT images, its performance, however, may be degraded in realistic patient cases with occasionally irregular breathing motions, such as cough. We plan to modify our current model to specifically take occasional irregularity into account and perform further tests in phantom data with realistic breathing motion and in real patient cases.

%Third, in practice, there remains other practical issues that one may encounter, e.g. the data truncation problem while imaging a large patient with a relatively small field of view and the contamination of scattering signals to the projections. All of these will be studied in our future work.

\section*{Acknowledgement}
Steve Jiang and Xun Jia would like to acknowledge the support from NIH (1R01CA154747-01, the Master Research Agreement from Varian Medical Systems, Inc., and the Early Career Award from Thrasher Research Fund. Hao Gao acknowledges the support from NIH/NIBIB (R21EB014956). Hongkai Zhao acknowledges the support from ONR (N00014-11-1-0602) and NSF (DMS-1115698 and 0928427).

\newpage
\section*{References}

\bibliographystyle{jmr}%
\bibliography{ref_jia,ref_cai}

\end{document}